\documentclass[10pt,twocolumn,letterpaper]{article}

\usepackage{cvpr}
\usepackage{times}
\usepackage{epsfig}
\usepackage{graphicx}
\usepackage{amsmath}
\usepackage{amssymb}
\usepackage{longtable}
\usepackage{paralist}
\usepackage{multirow,epstopdf,subfigure}
\usepackage{enumitem}
\usepackage{algorithm}
\usepackage{algorithmic}
\usepackage{bm}
\usepackage{color} 

\usepackage[breaklinks=true,bookmarks=false]{hyperref}

\cvprfinalcopy 

\def\cvprPaperID{****} 

\begin{document}

\title{Adaptive noise imitation for image denoising}

\author{Huangxing Lin\textsuperscript{1}, \ Yihong Zhuang\textsuperscript{1}, \  Yue Huang\textsuperscript{1}, \  Xinghao Ding\textsuperscript{1*},\ Yizhou Yu\textsuperscript{2}, \ Xiaoqing Liu\textsuperscript{3}, \ John Paisley\textsuperscript{4}\\
\normalsize \textsuperscript{1}Xiamen University,
\normalsize \textsuperscript{2}the University of Hong Kong,\\
\normalsize \textsuperscript{3}Deepwise,
\normalsize \textsuperscript{4}Columbia University\\
{\small \tt  $\{hxlin, zhuangyihong \}$@stu.xmu.edu.cn, huangyue05@gmail.com, dxh@xmu.edu.cn,}\\
{\small \tt yizhouy@acm.org, Liuxiaoqing@deepwise.com, jwp2128@columbia.edu}
}

\maketitle


\begin{abstract}
    The effectiveness of existing denoising algorithms typically relies on accurate pre-defined noise statistics or plenty of paired data, which limits their practicality. In this work, we focus on denoising in the more common case where noise statistics and paired data are unavailable. Considering that denoising CNNs require supervision, we develop a new \textbf{adaptive noise imitation (ADANI)} algorithm that can synthesize noisy data from naturally noisy images. To produce realistic noise, a noise generator takes unpaired noisy/clean images as input, where the noisy image is a guide for noise generation. By imposing explicit constraints on the type, level and gradient of noise, the output noise of ADANI will be similar to the guided noise, while keeping the original clean background of the image. Coupling the noisy data output from ADANI with the corresponding ground-truth, a denoising CNN is then trained in a fully-supervised manner. Experiments show that the noisy data produced by ADANI are visually and statistically similar to real ones so that the denoising CNN in our method is competitive to other networks trained with external paired data.

\end{abstract}

\section{Introduction}

Image denoising is an ill-posed inverse problem to recover a clean signal $y$ from the corrupted noisy image $x$,

\begin{equation}
\label{eq.x}
x=y+n,
\end{equation}
where $n$ is the noise component we would like to remove. In many imaging systems \cite{lee1999polarimetric, eo2018kiki}, image noise comes from multiple sources, such as the capturing instrument, medium of data transmission, and subsequent postprocessing. This complex generation process leads to complex noise distributions and variable noise levels, which makes denoising a challenging problem.

Recently, the field of image denoising has become dominated by supervised deep convolutional neural networks  (CNN), for which a noisy input and the corresponding ground-truth are required. Many CNNs \cite{zhang2018ffdnet, yue2019variational} show impressive denoising performance on some synthetic datasets. However, the synthesized noise usually deviates severely from the real noise distribution, resulting in often poor generalization. In addition, for many imaging systems, such as medical imaging, paired data is difficult to obtain, further limiting the application of these supervised techniques.

\begin{figure}
	\centering
	\includegraphics[width=0.73in]{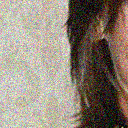}
	\includegraphics[width=0.73in]{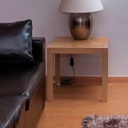}
	\includegraphics[width=0.73in]{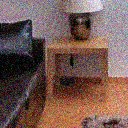}
	\includegraphics[width=0.73in]{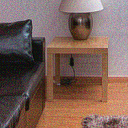}\\
	

	\includegraphics[width=0.73in]{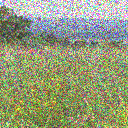}
	\includegraphics[width=0.73in]{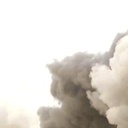}
	\includegraphics[width=0.73in]{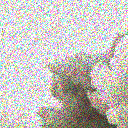}
	\includegraphics[width=0.73in]{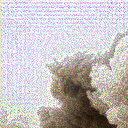}\\
	
	\vspace{-0.06in}

	\subfigure[Noise]{\includegraphics[width=0.73in]{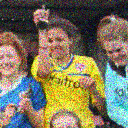}}
	\subfigure[Clean]{\includegraphics[width=0.73in]{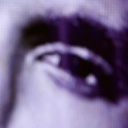}}
	\subfigure[ADANI]{\includegraphics[width=0.73in]{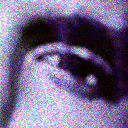}}
	\subfigure[LIR \cite{du2020learning}]{\includegraphics[width=0.73in]{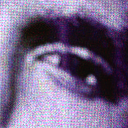}}

	\caption{(a) and (b): unpaired data for noise generation. (c): the image produced by ADANI for supervision has noise similar to (a), while the background is (b). (d): LIR is a GAN-based method that also uses unpaired data (a) and (b) to generate a noisy image. From top to bottom the noise are Gaussian, Speckle and Poisson.}
	\label{fig1}
\end{figure}

To relax data constraints, training denoising CNNs without pre-collected paired data has become a focus topic. Some ``self-supervised'' methods, such as Noise2Void \cite{krull2019noise2void} and Self2Self \cite{quan2020self2self}, show that individual noisy images can be used to train denoising networks via the so-called blind spot strategy. Despite great success, the effectiveness of these self-supervised methods relies on some pre-defined statistical assumptions, for example, noise $n$ is zero-mean, $n$ is independent of clean signal $y$, and $n$ is pixel-independent. Once the noise distribution does not meet those assumptions (\textit{e.g.} speckle noise), the denoising performance drops significantly. 

\begin{figure}
	\centering
	
	\subfigure[Gaussian noise]{\includegraphics[width=1.05in]{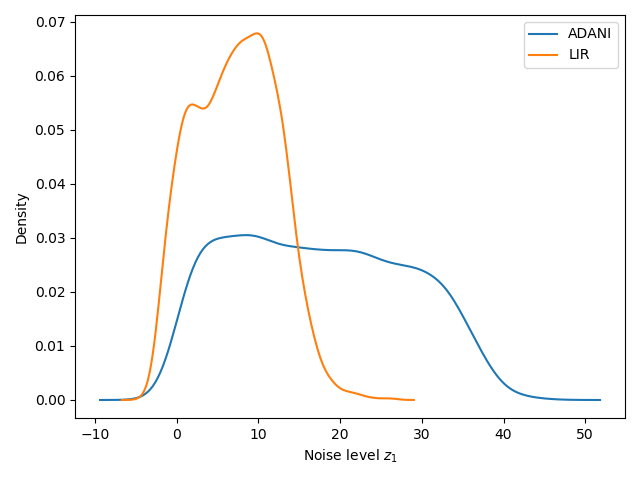}}
	\subfigure[Speckle noise]{\includegraphics[width=1.05in]{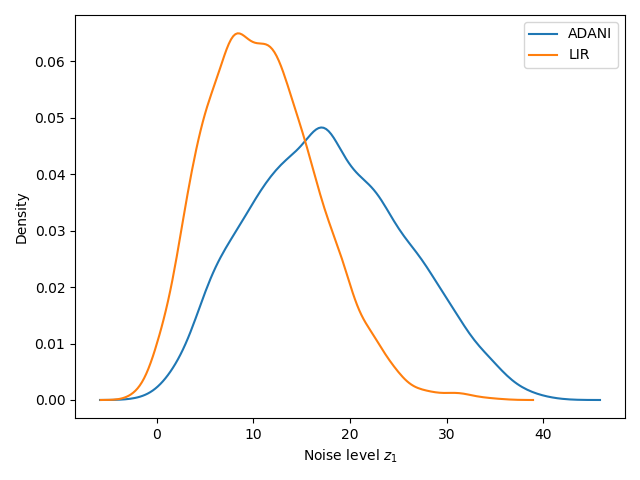}}
	\subfigure[Poisson noise]{\includegraphics[width=1.05in]{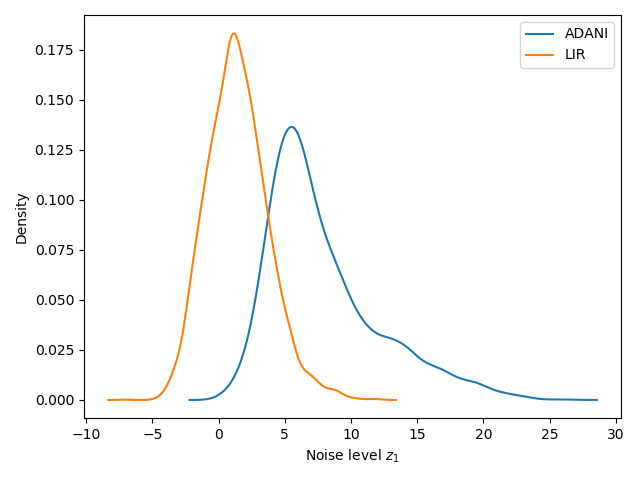}}
	
	\caption{The noise level statistical histograms for 10,000 images generated by ADANI and LIR. (a)-(c) represent the noisy data generated in three experiments. The level of noise produced by LIR is similar. In contrast, the output of ADANI has a wider distributional coverage. The noise level $z_1$ is provided by a pre-trained noise level estimator.}
	\label{fig_level}
\end{figure}

Another elegant strategy is to use unpaired noisy/clean images to learn denoising. 
To generate the necessary supervision, methods along this line usually integrate noise modeling and denoising into a deep learning framework. For instance, the authors in \cite{chen2018image, du2020learning, kaneko2020noise} use generative adversarial network (GAN) \cite{goodfellow2014generative} to synthesize noisy images corresponding to accessible clean images for supervision. Due to its strong generative ability, GAN is currently the most popular tool for unpaired denoising. However, GAN cannot promise the quality of the generated data, so the generated noise is often unrealistic (see Figure \ref{fig1}). In addition, GAN suffers from mode collapse \cite{arjovsky2017wasserstein}, resulting in a lack of diversity in the generated data. In Figure \ref{fig_level}, we see that the noisy images generated by the GAN-based method LIR exhibit monotonous noise levels. These unrealistic and monotonous noisy images will lead to poor denoising.

Motivated by the practical value of this open problem, we develop an efficient denoising method that does not rely on pre-defined noise statistics or pre-collected paired data. Given that the collection of unpaired data is relatively easy in most applications, the focus of our work is unpaired learning.  Similar to the previous methods \cite{chen2018image, du2020learning, kaneko2020noise, yan2019unsupervised}, our strategy is to use unpaired data to synthesize new noisy images to learn a denoising model. We also use the GAN as part of our model to distinguish the types of noise. However, the key of our method is to generate realistic noise with adjustable noise level by imitating a guided noise. In this way, we can simply change the guided noisy image to generate a variety of different levels of noise, thereby expanding the distributional coverage of noise.
Specifically, the generated noise is forced to be similar to a guided real noise by comparing their gradients, thus avoiding unrealistic noise patterns. Then, a pre-trained noise/clean classification network is introduced to estimate the level of noise. To achieve the same noise level, the noise generator is encouraged to imitate the guided noise to refine its output noise. For the background of the generated image, it is consistent with an accessible clean image specified by a background consistency module. Since the generator can adaptively generate noise similar to the input guided noise, we call our method adaptive noise imitation (ADANI) algorithm. Next, by pairing the generated noisy image with the corresponding ground-truth, we can train a denoising CNN supervisedly. To demonstrate the effectiveness of ADANI, we conduct experiments on several synthetic and real-world datasets. The noisy image produced by ADANI is visually and statistically indistinguishable from the real one. Consequently, the performance of our denoising CNN is close to other networks trained with pre-collected paired data and is better than other self-supervised and unpaired denoising methods.

Overall, our contributions are summarized as follows:

\begin{itemize}
	\item We propose an adaptive noise imitation algorithm for the generation of various noisy images, which only requires some unpaired data.
	
	\item We observe that the class logit (the input to the final softmax) from the noise/clean classification network is positively correlated with the noise level of the image. We use it as an indicator of the noise level.
	
	\item We show the application of the data generated by ADANI in various denoising tasks, where the noise statistics can be unknown.
	
\end{itemize}

\begin{figure*}
	\centering
	\includegraphics[width=5.6in]{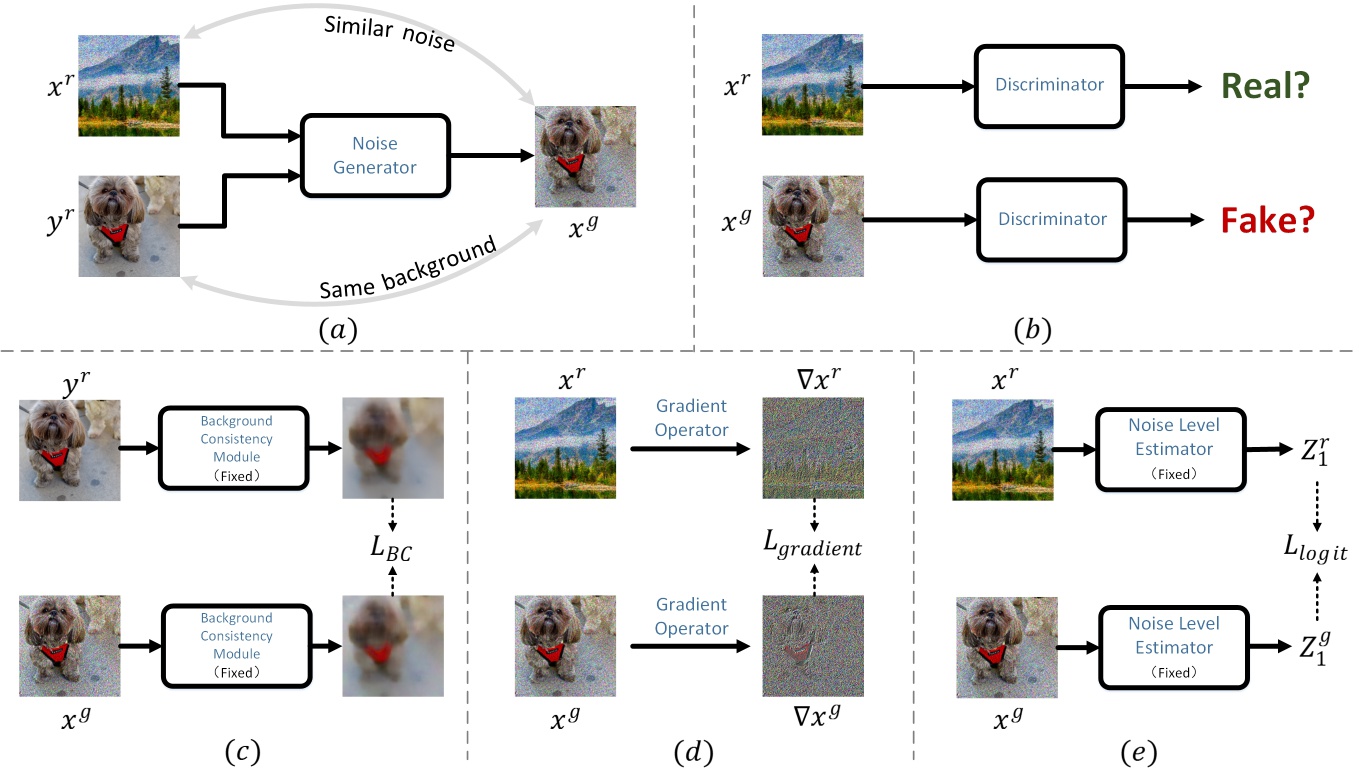}
	
	\caption{Illustration of our adaptive noise imitation algorithm. The generated $x^g$ has noise similar to $x^r$, and its background is $y^r$. 
	}
	\label{fig_frame}
\end{figure*}

\section{Related Work}
We present a brief review of image denoising methods related to our work, including model-based methods and learning-based methods.

\emph{\textbf{Model-based methods.}} Most traditional image denoising algorithms use hand-crafted priors \cite{xu2018trilateral, buades2005non, xu2018trilateral, meng2013robust, zhao2014robust} to simplify the denoising problem. One widely-used prior in image denoising is non-local self-similarity (NSS) \cite{hou2020nlh, mairal2009non, dong2012nonlocally}, which assumes that many patches in a non-local image area share a similar appearance. Some popular NSS-based algorithms, such as BM3D \cite{dabov2007image} and WNNM \cite{gu2014weighted}, have become the benchmark of image denoising. Other prominent techniques, such as total variation \cite{rudin1992nonlinear, beck2009fast, selesnick2017total}, wavelet coring \cite{simoncelli1996noise} and low-rank assumptions \cite{zhu2016noise, chang2017hyper}, have also been proven effective in some image denoising tasks. 

\emph{\textbf{Learning with paired data.}} Due to powerful nonlinear modeling capabilities, deep learning has become the dominant method for image denoising \cite{chang2020spatial, kim2020transfer, liu2020joint, zhang2020memory}.
Typically, supervised denoising CNNs require a large number of pairs of noisy/clean images for supervision. A state-of-the-art supervised approach is DnCNN \cite{zhang2017beyond}, which exploits residual learning for blind denoising. Following DnCNN, many different network architectures have been designed to obtain better visual results after denoising, including FFDNet \cite{zhang2018ffdnet}, DANet \cite{yue2020dual}, VDN \cite{yue2019variational}, RIDNet \cite{anwar2019real} and NLRN \cite{liu2018non}. 
When clean targets are unavailable, Lehtinen \emph{et al.} \cite{lehtinen2018noise2noise} suggest to learn a Noise2Noise (N2N) model with the pairs of two noisy images of the same scene. The performance of N2N is on par with other networks trained using noisy/clean image pairs. Nevertheless, it is not always feasible to sample two independent noises for the same scene. 

\emph{\textbf{Learning without paired data.}} It is sometimes useful to develop methods that do not rely on paired data as inputs. The blind-spot mechanism proposed in Noise2Void (N2V) \cite{krull2019noise2void} allows the denoiser to be trained using individual noisy images without paired data. This implementation is based on the assumption that the noise is zero mean and spatially uncorrelated, so that each pixel can be restored by its surrounding pixels. Due to its practical value, the blind spot mechanism is further improved in \cite{laine2019high, batson2019noise2self, wu2020unpaired, quan2020self2self} to obtain higher quality denoising. However, these self-supervised methods cannot handle noise that violates their assumptions, such as spatially correlated noise. In contrast, our ADANI does not rely on assumptions about the statistical characteristics or patterns of noise. Another strategy is to train denoising CNNs with unpaired noisy/clean images, which is also the focus of our work. Since unpaired data cannot directly guide the denoising for CNNs, methods to this group usually learn to synthesize noise before denoising. In particular, the GAN is widely used for noise modeling and has shown the potential for blind image denoising \cite{chen2018image,  kaneko2020noise, yan2019unsupervised}. Furthermore, cycle-consistency \cite{du2020learning} is utilized to aid GAN in learning the invariant representation between noise and clean domains. However, the discriminant information from GAN is too ambiguous to permit the generation of complex and diverse noise. Therefore, these GAN-based methods are easy to suffer from mode collapse, resulting in lack of diversity or unrealistic noise.

\section{Methodology}
Given some unpaired noisy images $\mathcal{D}^{noise}=\{x_i \}^N_{i=1}$ and clean images $\mathcal{D}^{clean}=\{y_j \}^M_{j=1}$, our goal is to learn denoising with these unpaired data. We denote the data distribution as $x^r \sim p^r(x)$ and $y^r \sim p^r(y)$. Hereafter, we use superscripts $r$ and $g$ to represent the real distribution and generative distribution, respectively. To reconstruct high-quality images, supervised methods \cite{zhang2017beyond, yue2019variational} incorporate pixel-level constraints to inform the denoising CNN to carefully restore each pixel during denoising. Unfortunately, unpaired data cannot directly form pixel-level supervision due to different image content. To solve this problem, we propose an adaptive noise imitation (ADANI) algorithm, which uses a CNN to learn to synthesize noise with these unpaired data (see Figure \ref{fig_frame}). In doing so, the gound-truth corresponding to the newly generated noisy image provides strong supervision for denoising. 

\subsection{GAN-based noise generation}
GANs have recently demonstrated the potential to generate certain distribution types of noise  \cite{chen2018image, du2020learning, kaneko2020noise}. We also adopt GAN as a component of our method to guide noise generation. The process of noise generation is performed by a generator that takes a clean background image $y^r$ and a guided noisy image $x^r$ as input,

\begin{equation}
\label{eq.g}
\begin{aligned}
x^g&=G(y^r,x^r)\\
&=y^r+n^g.
\end{aligned}
\end{equation}
where $x^g \sim p^g(x)$ and $n^g$ is the noise generated by the generator. We use the extra input $x^r$ to guide the generation of realistic noise.

\emph{\textbf{Background consistency.}} In Eq.(\ref{eq.g}), we want to generate noisy $x^g$ with the same image background as $y^r$, so that $x^g$ and $y^r$ can be paired to train denoising CNNs. To this end, we first build a background consistency module (BCM) to preserve the background consistency between $x^g$ and $y^r$. BCM is a pre-trained network related to image filters (\textit{e.g.} Gaussian filter, median filter). It is based on the assumption that paired noisy and clean images share the same low-frequency content. To pre-train BCM, a mixture of $\mathcal{D}^{noise}$ and $\mathcal{D}^{clean}$ is adopted as the training set $\mathcal{D}^{mix}$, and the blurred targets corresponding to $\mathcal{D}^{mix}$ is produced by image filtering (we use a median filter with a kernel size of 31). After pre-training, BCM acts as an image filter, which can filter out high-frequency parts including noise from the input image. We use BCM to provide the background consistency constraint,

\begin{equation}
\label{eq.Lbc}
\mathcal{L}_{BC}=\mathbb{E}_{y^r \sim p^r(y),\hfill\atop x^g \sim p^g(x) \hfill} \left[ \left\| {{B(x^g)} - B(y^r)} \right\|_1 \right],
\end{equation}
where $B(\cdot)$ denotes BCM and we adopt L1 loss.

To generate noise, the generator in Eq.(\ref{eq.g}) is supervised by a noise discriminator. Following adversarial training, a discriminator is responsible for distinguishing a real noisy image $x^r$ from the generated noisy image $x^g$. The purpose of the generator is to fool the discriminator which means that $p^g(x)$ gets close to $p^r(x)$. This corresponds to the following GAN loss,

\begin{equation}
\scriptsize
\label{eq.gan}
\mathcal{L}_{GAN}=\mathbb{E}_{x^r \sim p^r(x)} \left[log D(x^r)\right]+\mathbb{E}_{x^g \sim p^g(x)} \left[log \left(1-D(x^g) \right) \right].
\end{equation}

Eq.(\ref{eq.gan}) allows the generation of noise of a certain distribution type, but does not impose constraints on the quality of noise. Therefore, unrealistic noise is often generated. In addition, Eq.(\ref{eq.gan}) does not indicate the level of noise, which leads to mode collapse. For example, use Eq.(\ref{eq.g}) to synthesize Gaussian noise with different variances (\textit{e.g.} $\sigma \in \left( 0, 50 \right]$). The generator can easily fool the discriminator by always producing the same level (\textit{e.g.} $\sigma = 25$) of noise, resulting in the lack of variety.

To solve the above problems, our strategy is to generate $x^g$ similar to the guided noise $x^r$ in noise type and level by imitating $x^r$. Since the noise of $x^g$ is similar to $x^r$, it avoids unrealistic noise patterns. What's more, we can obtain various noisy data by simply changing $x^r$. We then introduce some constraints on the noise similarity between $x^r$ and $x^g$.

\begin{figure}
	\centering

	\includegraphics[width=0.78in]{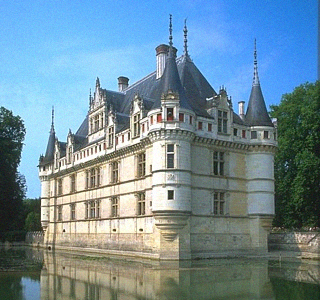}
	\includegraphics[width=0.78in]{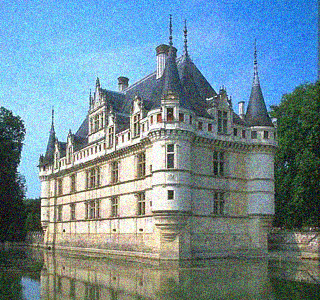}
	\includegraphics[width=0.78in]{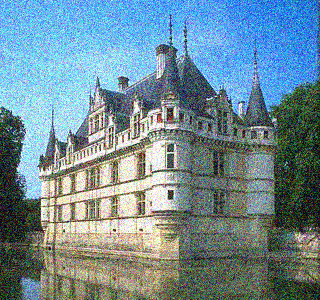}
	\includegraphics[width=0.78in]{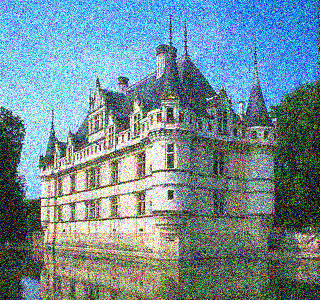}\\
	\vspace{-0.04in}

	\subfigure[{$\sigma=5$, \color{white}{aa} \color{black} $ z_1=1.09$,\color{white}{aaa} \color{black} $q_1=0.91$.}]{\includegraphics[width=0.78in]{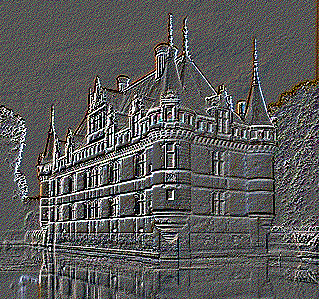}}
	\subfigure[$\sigma=20$, \color{white}{aa} \color{black} $ z_1=10.6$,\color{white}{aaa} \color{black} $q_1=1.00$.]{\includegraphics[width=0.78in]{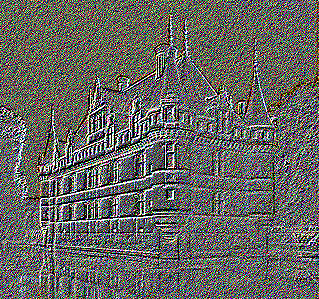}}
	\subfigure[$\sigma=35$, \color{white}{aa} \color{black} $ z_1=21.6$,\color{white}{aaa} \color{black} $q_1=1.00$.]{\includegraphics[width=0.78in]{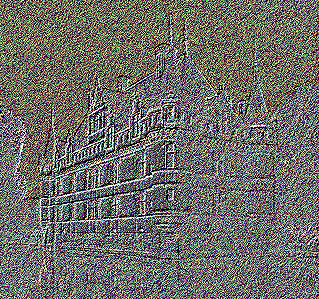}}
	\subfigure[$\sigma=50$, \color{white}{aa} \color{black} $ z_1=30.0$,\color{white}{aaa} \color{black} $q_1=1.00$.]{\includegraphics[width=0.78in]{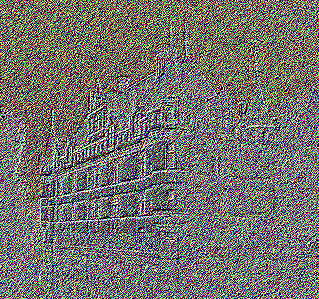}}
	
	\caption{Images with different levels of Gaussian noise. In the second row are gradient maps corresponding to the first row. $\sigma$: the standard deviation of Gaussian noise. $z_1$ and $q_1$: the logit and probability output by the noise level estimator Eq.(\ref{eq.softmax}). }
	\label{fig_house}
\end{figure}

\emph{\textbf{Noise similarity.}} We note that image noise is a random variation in pixel brightness, which will cause the magnitude of the gradient around the noisy pixel to be dramatically improved. The image gradient reflects the high-frequency information (\textit{e.g.} noise, edges), while excluding the low-frequency image content. Normally, the more noisy the image is, the more noisy the gradient map is. This motivates us to achieve noise similarity by matching the gradient distributions of $x^r$ and $x^g$. We compute the image gradient $\nabla x$ by combining the horizontal and vertical deviations of adjacent pixels. Then, we impose an L1 penalty on the gradient gap between $x^r$ and $x^g$,

\begin{equation}
\begin{aligned}
\label{eq.Lgrad}
\mathcal{L}_{gradient}=&\mathbb{E}_{x^r \sim p^r(x),\hfill\atop x^g \sim p^g(x) \hfill}  [\left\| { {{\nabla }x^g}-{{\nabla }x^r}} \right\|_1 ].
\end{aligned}
\end{equation}

Since the gradient of the noisy image is dominated by noise, $\mathcal{L}_{gradient}$ forces the noise of $x^g$ to be similar to the real $x^r$.
Combining Eq.(\ref{eq.Lbc}), Eq.(\ref{eq.gan}) and Eq.(\ref{eq.Lgrad}), our GAN-based noise generation strategy can be briefly expressed as

\begin{equation}
\label{eq.L_all}
\mathop {\min }\limits_G \mathop {\max }\limits_D {\mathcal{L}_{GAN}} + \alpha \mathcal{L}_{gradient}+ \beta \mathcal{L}_{BC},
\end{equation}
where $\alpha$ and $\beta$ are the trade-off parameters.

\subsection{Adaptive noise imitation}
In Eq.(\ref{eq.L_all}), $\mathcal{L}_{GAN}$, $\mathcal{L}_{gradient}$ and $\mathcal{L}_{BC}$ are the constraints on noise type, gradient and image background, respectively. Among them, $\mathcal{L}_{gradient}$ can be regarded as an indicator of the similarity between the generated noise and the guided noise. However, a flaw of $\mathcal{L}_{gradient}$ is that it gives equal importance to guided noisy images over a wide range of noise levels. For the guided $x^r$ with a high level of noise, the primary component of its gradient map is noise, which can provide effective guidance for noise generation. On the contrary, for $x^r$ with weak noise, its gradient map is mainly composed of edges (see Figure \ref{fig_house}), which may pollute the generated noise. These observations suggest that different $x^r$ has different effects on noise generation. Therefore, we set the hyperparameter $\alpha$ for $L_{gradient}$ Eq.(\ref{eq.L_all}) to adaptively change according to the noise level of $x^r$, rather than a fixed value. The noisier the $x^r$ is, the greater the $\alpha$ is, so as to eliminate the negative effect of $L_{gradient}$. To do this, we use a pre-trained noise/clean binary classification network as a noise level estimator, where the noisy image is class 1 and the clean image is class 0. The dataset for pre-training the classification network is also $D^{mix}$. Class probabilities are produced by the softmax activation layer that converts the logit, $z_i$, computed for each class $i$ into a probability, $q_i$, by comparing $z_i$ with the other logits,

\begin{equation}
\begin{aligned}
\label{eq.softmax}
&{q_i} = \frac{{\exp ({z_i})}}{{\sum\nolimits_j {\exp ({z_j})} }},\\
&z_i=C(x)_i,
\end{aligned}
\end{equation}
where $j\in \{0,1 \}$, $C(\cdot)$ denotes the classification network except the softmax layer. After pre-training, the classification probability $q_i$ represents the network's confidence that its input belongs to class $i$. This means that for noise/clean classification, $q_1$ reflects the noise level to a certain extent. However, $q_1 \in [0,1]$, it is difficult to cover a wide range of noise. In addition, the early saturation behavior \cite{chen2017noisy} of softmax makes most noisy images easily classified into class 1 with high confidence (\textit{i.e.} $q_1 \rightarrow 1$). Therefore, $q_1$ produced by softmax cannot accurately characterize the noise level. 

Based on the above analysis, when learning to generate noise, we remove the softmax layer from the noise level estimator and use the logit $z_1$ as the estimate of the noise level. The value of $z_1$ is not limited, so it can match a wide range of noise. More importantly, $z_1$ is positively correlated with the noise level of input, as shown in Figure \ref{fig_house}. 
Therefore, in each iteration, the level of guided noise is estimated by,

\begin{equation}
\label{eq.zr}
{z^r_1} = C(x^r)_1.
\end{equation}

Then, we use $z^r_1$ to replace $\alpha$ in Eq.(\ref{eq.L_all}). Following such a dynamic objective, the noise generator can identify the components of interest (\textit{i.e.} noise) in the guided $x^r$, and adaptively generate realistic noise similar to $x^r$. Finally, we construct a logit consistency loss to further promote the noise similarity between $x^g$ and $x^r$, \textit{i.e.}

\begin{equation}
\begin{aligned}
\label{eq.Llogit}
\mathcal{L}_{logit}=&\mathbb{E}_{x^r \sim p^r(x),\hfill\atop x^g \sim p^g(x) \hfill}  [\left\| z^g_1-z^r_1 \right\|_2 ],
\end{aligned}
\end{equation}
where ${z^g_1} = C(x^g)_1$.

Combining Eq.(\ref{eq.L_all}), Eq.(\ref{eq.Llogit}) and ${z^r_1}$, we aim to solve,

\begin{equation}
\label{eq.L_all_final}
\mathop {\min }\limits_G \mathop {\max }\limits_D {\mathcal{L}_{GAN}} + \alpha \mathcal{L}_{gradient}+ \beta \mathcal{L}_{BC}+ \gamma \mathcal{L}_{logit},
\end{equation}
where $\alpha=z^r_1$, $\beta$ and $\gamma$ are the trade-off parameters.

\begin{figure*}[t]
	\centering
	
	\subfigure[ {Clean $|$ SSIM, PSNR}]{\includegraphics[width=1.6in]{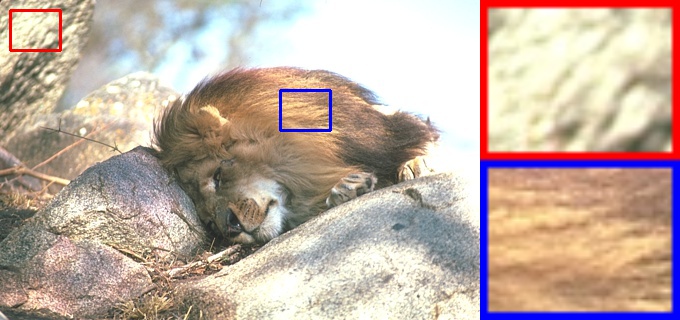}}
	\subfigure[ {Input $|$ 0.637, 24.06}]{\includegraphics[width=1.6in]{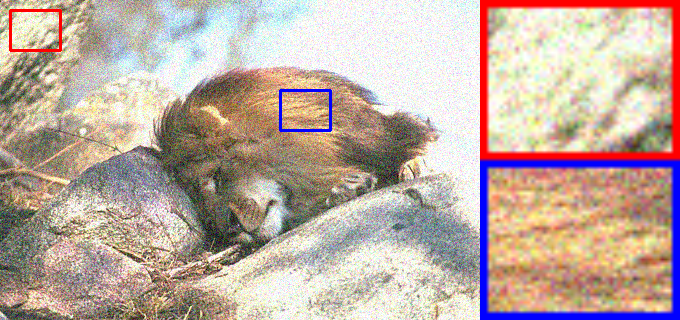}}
	\subfigure[ {BM3D $|$ 0.880, 28.56}]{\includegraphics[width=1.6in]{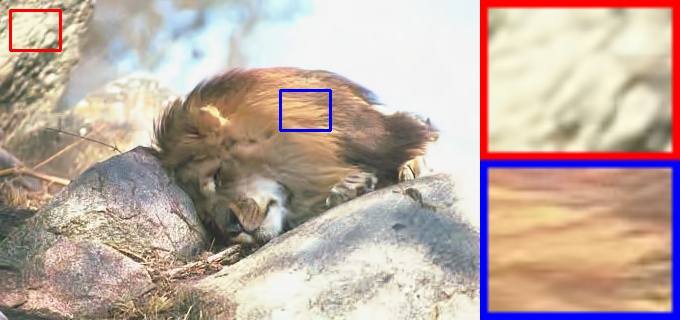}}
	\subfigure[ {N2V $|$ {0.889}, {28.43}}]{\includegraphics[width=1.6in]{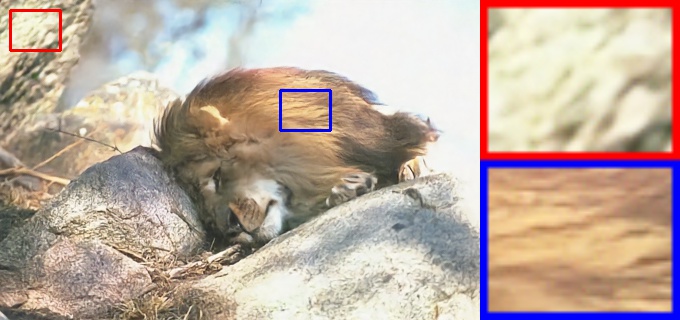}}
	\\
	\vspace{-0.1in}
	\subfigure[ {S2S $|$ 0.813, 26.69}]{\includegraphics[width=1.6in]{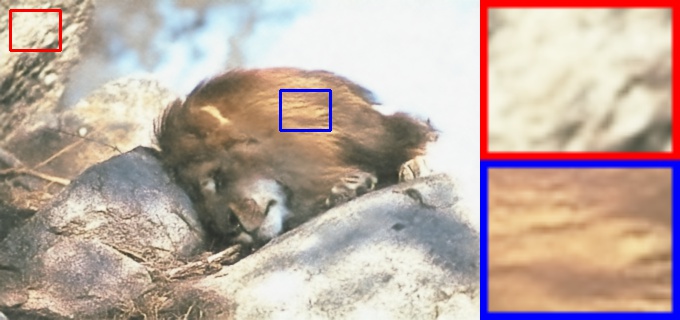}}
	\subfigure[ {LIR $|$ 0.867, 25.60}]{\includegraphics[width=1.6in]{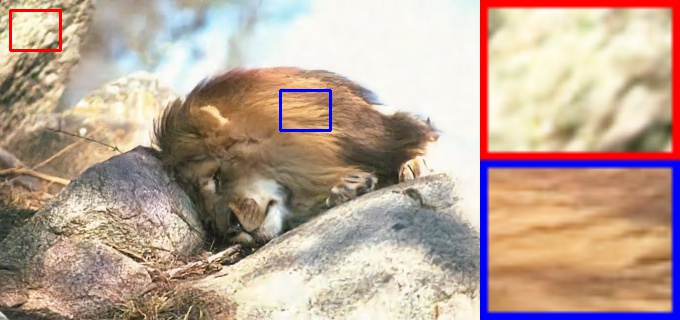}}
	\subfigure[ {U-Net $|$ \textbf{0.900}, \textbf{30.01}}]{\includegraphics[width=1.6in]{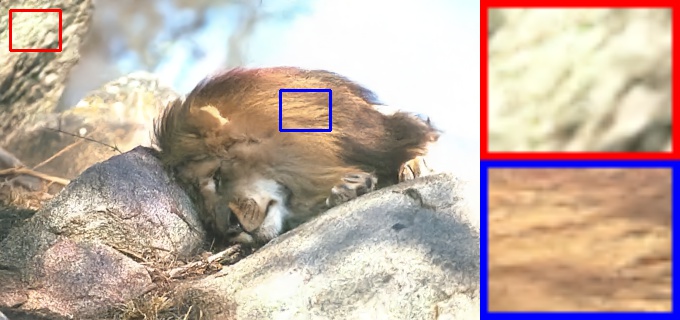}}
	\subfigure[ {Our $|$ 0.889, 29.38}]{\includegraphics[width=1.6in]{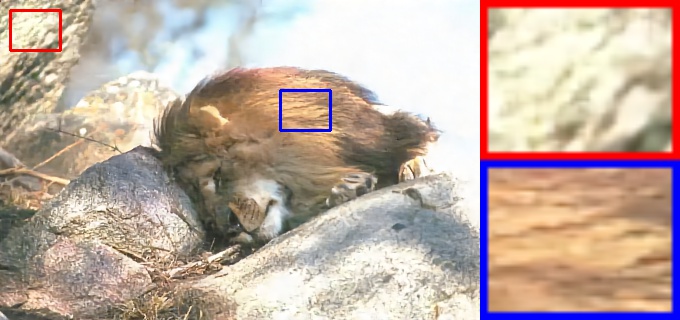}}
	
	\caption{Example results for Gaussian denoising, $\sigma=25$.}
	\label{fig_gaus}
\end{figure*} 

\begin{table*}[t]
	\caption{PSNR results (dB) from BSD300 dataset for Gaussian, Speckle and Poisson noise. \textbf{Bold}: best. \textcolor{red}{Red}: second. \textcolor{blue}{Blue}: third.}
	\centering
	\begin{tabular}{|c|c|c|c|c|c|c|c|c|c|}
		\hline
		&Test noise level& BM3D & WNNM&N2V&S2S&LIR&N2N&U-Net&ADANI\\
		\hline
		\multirow{2}*{Gaussian }&$\sigma=25$&\textcolor{blue}{30.90}&29.96&{30.51}&29.13&{26.91}&\textcolor{red}{31.32}&\textbf{31.45}&30.68\\
		\cline{2-10}
		&$\sigma \in (0, 50]$&27.89&31.16&{31.67}&27.06&{26.38}&\textcolor{red}{32.82}&\textbf{33.14}&\textcolor{blue}{31.85}\\
		\hline
		\multirow{2}*{Speckle }&$v=0.1$&26.64&25.13&{28.40}&27.41&{25.66}&\textcolor{red}{31.12}&\textbf{31.18}&\textcolor{blue}{29.96}\\
		\cline{2-10}
		&$v \in (0, 0.2]$&26.70&25.39&{28.77}&27.23&{25.44}&\textcolor{red}{31.50}&\textbf{31.55}&\textcolor{blue}{30.34}\\
		\hline
		\multirow{2}*{Poisson }&$\lambda =30$&27.70&28.09&29.70&28.75 &{26.15}&\textcolor{red}{30.44}&\textbf{30.81}&\textcolor{blue}{29.85}\\
		\cline{2-10}
		&$\lambda \in [5, 50]$&27.23&27.36&28.72&27.71&{25.62}&\textcolor{red}{29.65}& \textbf{30.14}&\textcolor{blue}{28.87} \\
		\hline
		
	\end{tabular}
	\label{table1}
\end{table*}

\subsection{Architecture and training details}

The implementation of ADANI is based on CNNs. For simplicity, both noise generator and BCM adopt the ResNet \cite{he2016identity} architecture. The discriminator is a general “PatchGAN” classifier \cite{li2016precomputed, isola2017image}. The noise level estimator is a simple four-layer network. 

\textit{Pre-training.} The BCM and noise level estimator are pre-trained with $D^{mix}$, and their weights are fixed when learning noise generation. The input $128\times128$ patches are randomly cropped from the training set, and the training ends at the 200th epoch. We use Adam with a batch size of 1 to train networks. The learning rate is initialized to 0.0002 and is linearly decayed to 0 over the training process.

\textit{Training for noise generation and denoising.} The unpaired patches $x^r$ and $y^r$ are randomly cropped from $D^{noise}$ and $D^{clean}$. The hyper-parameters in Eq.(\ref{eq.L_all_final}) are set to $\beta=300$, $\gamma=0.1$. In each iteration, the noise generator produces a pair of data ($x^g$, $y^r$), which are directly used to guide an U-Net\footnote{More details for network architectures are shown in the supplement.} \cite{ronneberger2015u} to learn denoising according to L1 loss. Training ends at the 1000th epoch. Other parameters are the same as those of pre-training.

\section{Experiments}

In this section, we evaluate the performance of ADANI on several denoising tasks.

\subsection{Synthetic noises}
\label{synthetic}

To prepare the unpaired training data, we use the 4744 clean images provided in \cite{ma2016waterloo} to synthesize noisy images (\textit{i.e.} $D^{noise}$) with Matlab. Besides, we collect another 5000 clean images from Internet as the clean set $D^{clean}$. The compared methods are state-of-the-art model-based methods BM3D \cite{dabov2007image} and WNNM \cite{gu2014weighted}, self-learning methods Noise2Void(N2V) \cite{krull2019noise2void} and Self2Self(S2S) \cite{quan2020self2self}, an unpaired learning method LIR \cite{du2020learning}, other deep learning methods include Noise2Noise(N2N) \cite{lehtinen2018noise2noise} and a common fully-supervised U-Net. For fair comparison, N2N, U-Net and our ADANI adopt the same architecture to perform denoising. For BM3D, we set its hyperparameter to $\sigma=25$ when removing Gaussian noise with a standard deviation of 25, while in other cases, BM3D keeps the default settings (\textit{i.e.} $\sigma = 50$). Our test set is the widely used BSD300 \cite{martin2001database}.

\begin{figure*}[t]
	\centering
	
	\subfigure[ {Clean $|$ SSIM, PSNR}]{\includegraphics[width=1.5in]{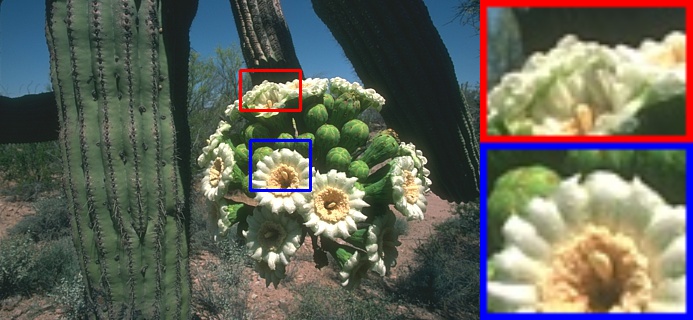}}
	\subfigure[ {Input $|$ 0.585, 22.11}]{\includegraphics[width=1.5in]{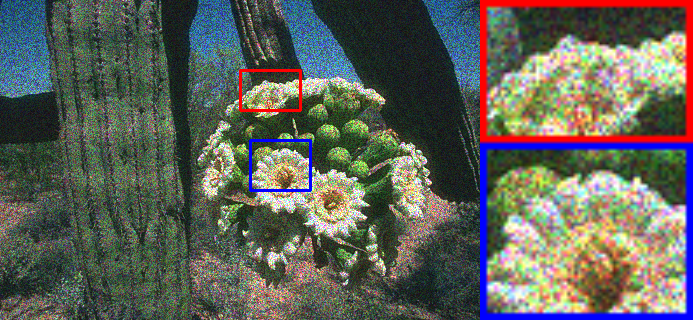}}
	\subfigure[ {BM3D $|$ 0.742, 25.98}]{\includegraphics[width=1.5in]{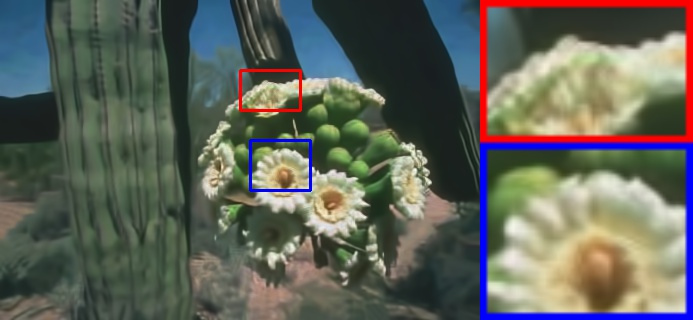}}
	\subfigure[ {N2V $|$ {0.868}, {28.00}}]{\includegraphics[width=1.5in]{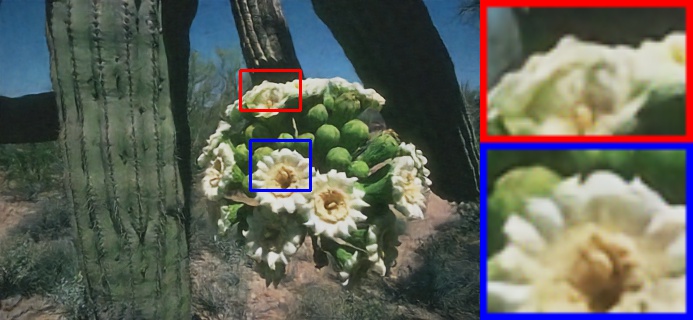}}
	\\
	\vspace{-0.1in}
	\subfigure[ {S2S $|$ 0.810, 26.80}]{\includegraphics[width=1.5in]{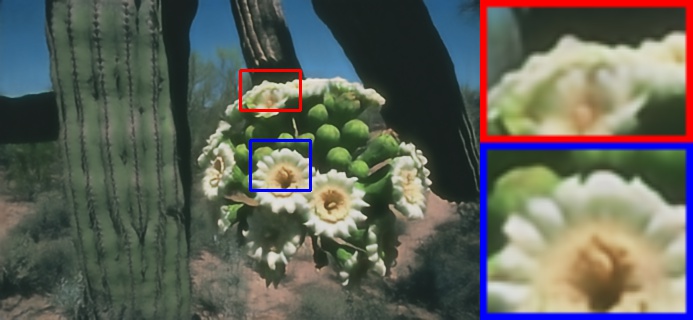}}
	\subfigure[ {LIR $|$ 0.835, 25.83}]{\includegraphics[width=1.5in]{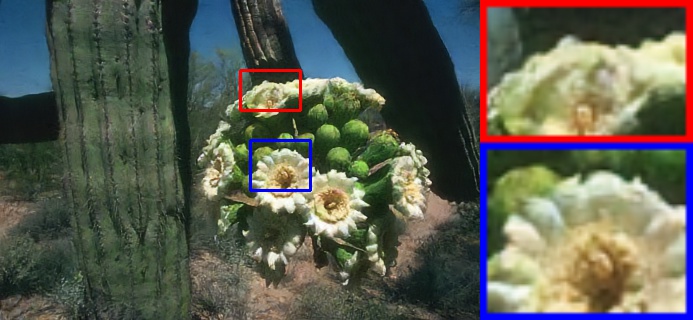}}
	\subfigure[ {U-Net $|$ \textbf{0.907}, \textbf{29.97}}]{\includegraphics[width=1.5in]{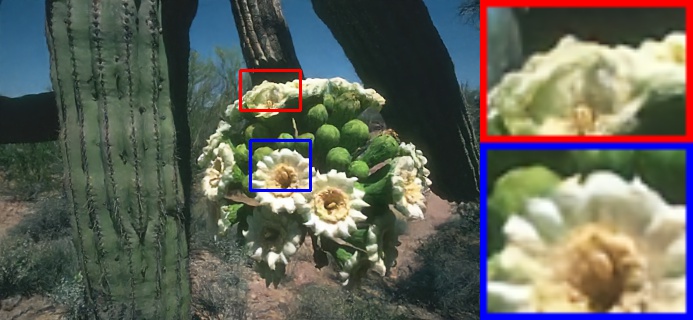}}
	\subfigure[ {Our $|$ 0.881, 29.16}]{\includegraphics[width=1.5in]{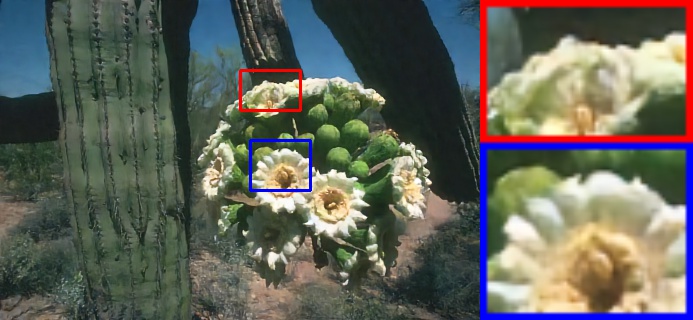}}
	
	\caption{Example results for Speckle denoising, $v=0.1$.}
	\label{fig_spk}
\end{figure*} 

\begin{figure*}[t]
	\centering
	
	\subfigure[ {Clean $|$ SSIM, PSNR}]{\includegraphics[width=1.5in]{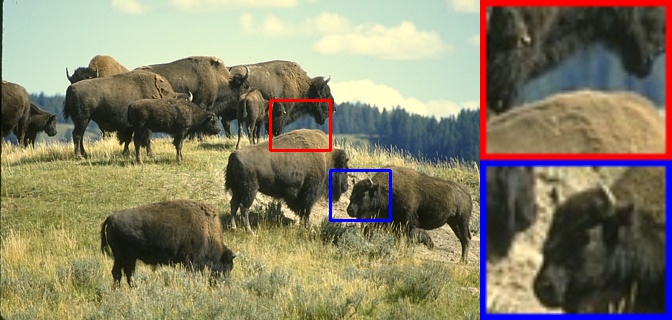}}
	\subfigure[ {Input $|$ 0.508, 21.42}]{\includegraphics[width=1.5in]{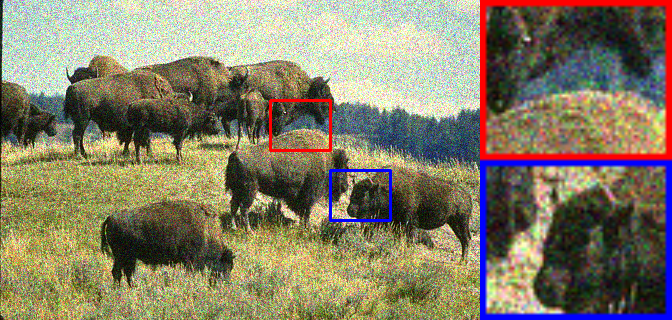}}
	\subfigure[ {BM3D $|$ 0.721, 25.45}]{\includegraphics[width=1.5in]{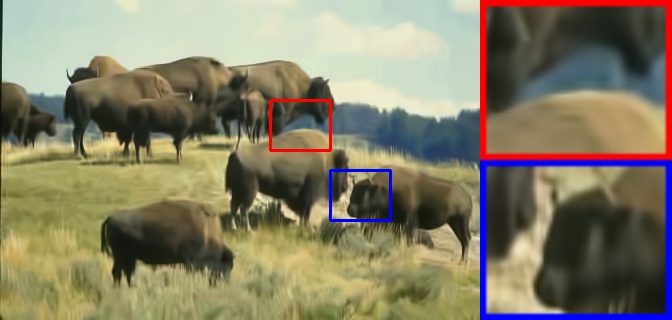}}
	\subfigure[ {N2V $|$ {0.847}, {27.56}}]{\includegraphics[width=1.5in]{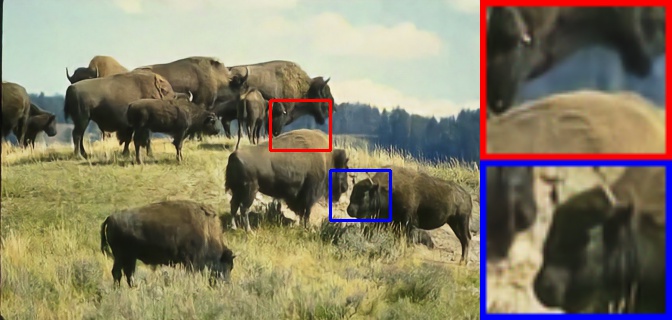}}
	\\
	\vspace{-0.1in}
	\subfigure[ {S2S $|$ 0.773, 26.23}]{\includegraphics[width=1.5in]{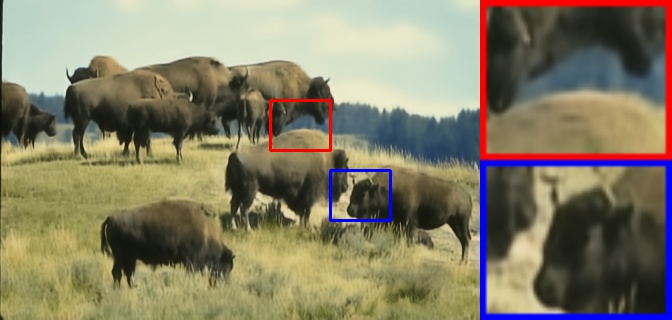}}
	\subfigure[ {LIR $|$ 0.810, 27.38}]{\includegraphics[width=1.5in]{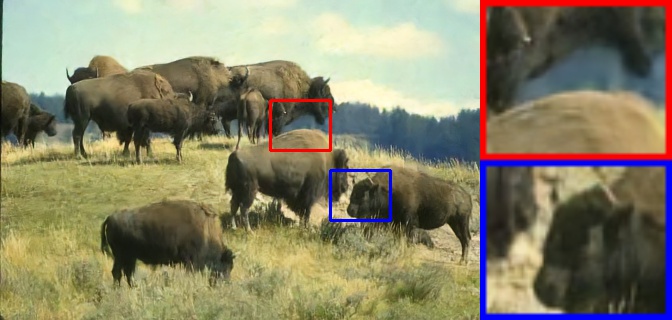}}
	\subfigure[ {U-Net $|$ \textbf{0.864}, \textbf{28.63}}]{\includegraphics[width=1.5in]{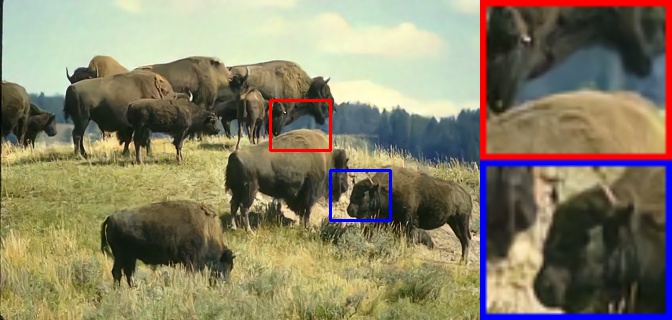}}
	\subfigure[ {Our $|$ 0.861, 27.78}]{\includegraphics[width=1.5in]{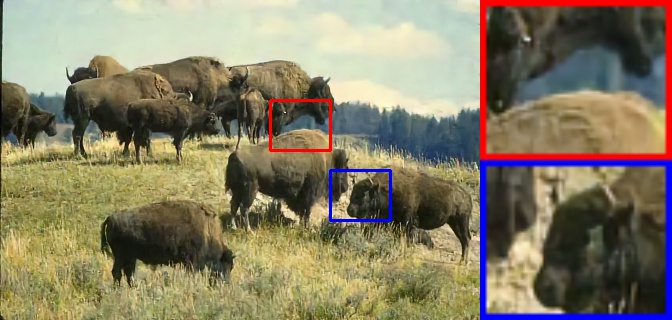}}
	
	\caption{Example results for Poisson denoising, $\lambda=30$.}
	\label{fig_pos}
\end{figure*} 

\begin{figure}
	\centering

	\subfigure[Gaussian $\sigma = 25$]{\includegraphics[width=1.05in]{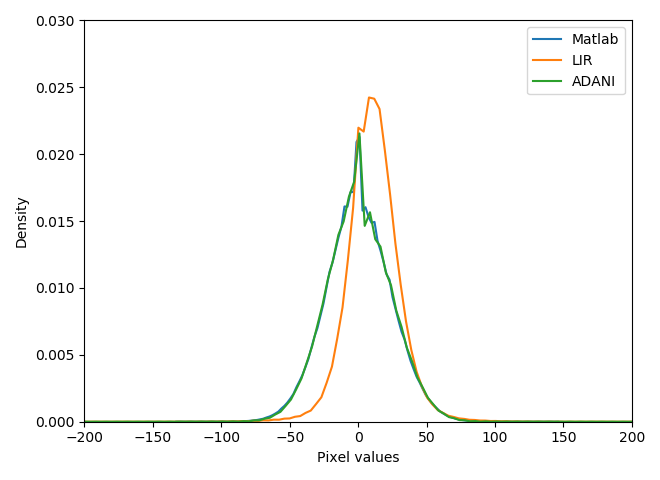}}
	\subfigure[Speckle $v=0.1$]{\includegraphics[width=1.05in]{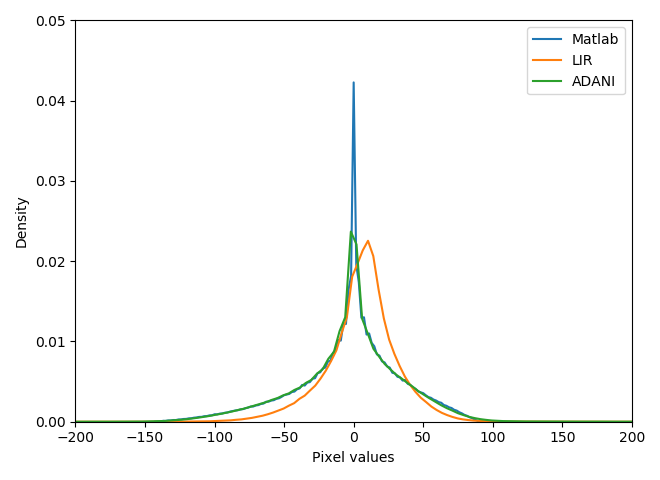}}
	\subfigure[Poisson $\lambda = 30$]{\includegraphics[width=1.05in]{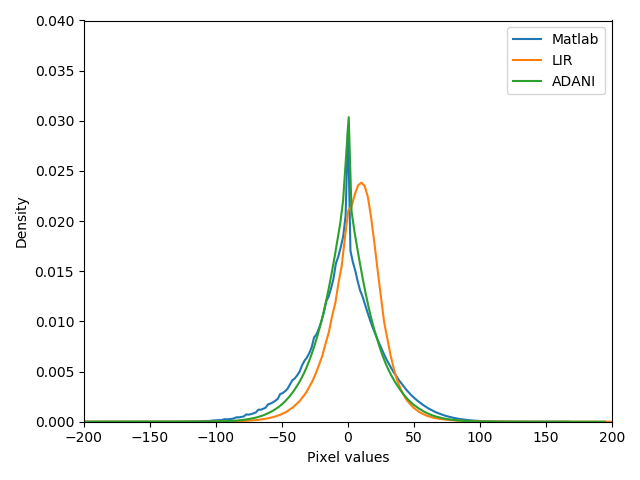}}
	
	\caption{Statistical histograms for noise generated by Matlab, ADANI and LIR. }
	\label{fig_hist}
\end{figure}

\begin{table*}[t]
	\caption{Quantitative results on $SIDD$ benchmark dataset. (CBDNet, VDN and U-Net are fully-supervised networks.)}
	\centering
	\begin{tabular}{|c|c|c|c|c|c|c|c|c|c|c|c|}
		\hline
		{} & {BM3D } &{WNNM}&{NLM}&{KSVD\cite{aharon2006k}}& {EPLL\cite{zoran2011learning}}& {CBDNet\cite{guo2019toward}} &VDN \cite{yue2019variational}&N2V&U-Net&    {ADANI}\\
		\hline
		
		{SSIM}&{0.685}&0.809&0.699&0.842&0.870&0.868&\textbf{0.955}&0.507&0.951&0.944\\
		
		\hline
		{PSNR}&{25.65}&25.78&26.75&26.88&27.11&33.28&\textbf{39.26}&22.41&38.68&37.64\\
		
		\hline
		
	\end{tabular}
	\label{table_real}
\end{table*}

\textbf{\textit{Gaussian noise. }} The first experiment is blind Gaussian denoising. Each training image is corrupted by Gaussian noise with a random standard deviation $\sigma \in \left( 0,50\right]$. For testing, we synthesize noisy images according to two strategies: a fixed noise level $\sigma = 25$ and a variable $\sigma \in \left( 0,50\right]$. Quantitative results are shown in Table \ref{table1}. Our method significantly outperforms other unpaired or self-learning methods, and is close to supervised networks (U-Net and N2N). Although our method is inferior to BM3D on the test set of Gaussian noise $\sigma = 25$, the effectiveness of BM3D relies on accurate noise priors. For noise with unknown distribution, the performance of BM3D is poor. In contrast, our method can be adapted to various noises. Figure \ref{fig_gaus} shows the denoising results of different competing methods. Our denoising network achieves promising results in removing noise and enhancing image quality.

\textbf{\textit{Speckle noise. }} To demonstrate the wide applicability of ADANI, we conduct experiments on speckle noise. Speckle noise is mostly detected in case of medical images and radar images. It is typically known as a multiplicative noise to the latent singal $y$, which can be modeled via the equation $x=y+y\cdot n$. In this equation, $n$ is the noise sampled from a uniform distribution with a mean of 0 and a variance of $v$. The noisy images for training are synthesized by varying the noise variance $v \in \left(0,0.2 \right]$. We report the comparison results in Table \ref{table1} and Figure \ref{fig_spk}. As can be seen, our ADANI consistently shows encouraging performance.

\textbf{\textit{Poisson noise. }} Poisson noise is usually used to model the photon noise of imaging sensors.  Its expected magnitude is signal dependent, so it is harder to remove than signal-independent noise. Following the setting in \cite{laine2019high}, we vary the noise magnitude $\lambda \in \left[5,50\right]$ during training.
Comparison results are presented in Table \ref{table1} and Figure \ref{fig_pos}.

\textbf{\textit{Discussion.}} These experiments on synthetic noise show the effectiveness and wide applicability of ADANI. It can generate realistic noisy images, as previously shown in Figure \ref{fig1}, to learn denoising, and the denoising performance is close to other networks trained with external paired data (U-Net and N2N). For practical applications where paired data is not available and noise statistics are unknown, our ADANI is better able to adapt than supervised methods.

\begin{figure}[t]
	\centering

	\subfigure[ {Noise}]{\includegraphics[width=1.3in]{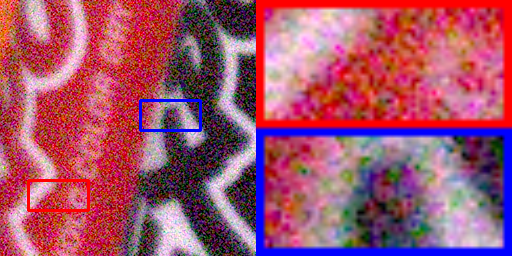}}
	\subfigure[ {N2V \cite{krull2019noise2void}}]{\includegraphics[width=1.3in]{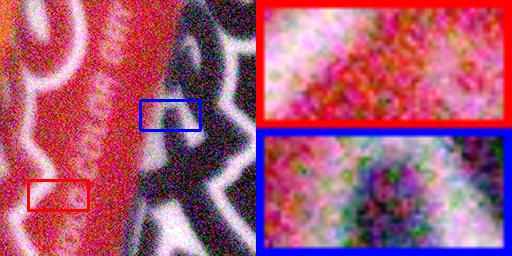}}\\
	\vspace{-0.1in}
	\subfigure[ {U-Net}]{\includegraphics[width=1.3in]{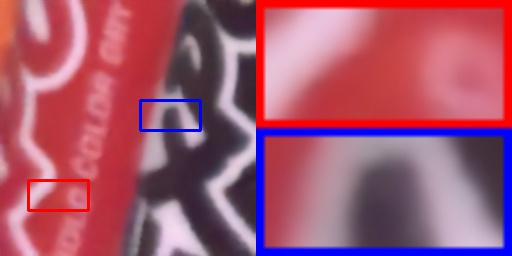}}
	\subfigure[ {Our }]{\includegraphics[width=1.3in]{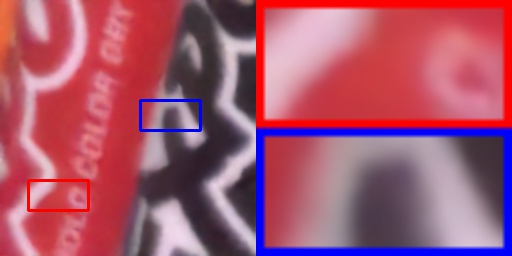}}
	
	\caption{Denoising results for SIDD dataset.}
	\label{fig_sidd}
\end{figure} 

\textbf{\textit{Noise statistics.}} We then demonstrate the ability of ADANI to model noise over a wide range of distribution. We randomly crop 10,000 image patches with a size of $128\times 128$ from $D^{clean}$. These clean image patches are input into the above three noise generators together with noisy patches randomly sampled from $D^{noise}$. The noise level estimates of the output of the generators are provided by the corresponding noise level estimators. Since LIR \cite{du2020learning} also uses unpaired data to generate noisy images, we compare ADANI with LIR. The noise level statistical histograms are shown in Figure \ref{fig_level}. As observed, the distributional coverage of data output by ADANI is much wider than that of LIR. This shows that our adaptive noise imitation strategy can avoid mode collapse.

We further evaluated the quality of noise generated by ADANI and LIR. To do this, we use the above 10,000 clean image patches to synthesize three noisy datasets with Matlab (\textit{i.e.} gaussian noise $\sigma =25$, speckle noise $v=0.1$, poisson noise $\lambda =30$). These noisy and clean image patches are randomly shuffled to form unpaired inputs for ADANI and LIR. To eliminate the influence of the background, the image background is subtracted from the noisy patch to obtain the noise component. Figure \ref{fig_hist} shows statistical histograms of noise generated by the three methods. As observed, the noise distribution produced by ADANI is similar to that of guided noise (Matlab), while the noise generated by LIR is obviously distorted. This experiment further demonstrates that ADANI can produce realistic noise by noise imitation.


\subsection{Real-world noise}
In this part, we evaluate the performance of ADANI on a real-world noise dataset Smartphone Image Denoising Dataset (SIDD) \cite{abdelhamed2018high}. SIDD contains thousands of images with various noise levels, dynamic ranges, and brightnesses. For each noisy image, the ground-truth is obtained via some statistical methods \cite{abdelhamed2018high}. For fast training, 320 pairs of high-resolution noisy/clean images are selected as the medium version of SIDD, called SIDD-Medium. We employ the SIDD-Medium dataset to train CNNs. To prepare unpaired data, SIDD-Medium is randomly divided into 2 parts, each with 160 pairs of images. We use 160 noisy images from the first part and 160 clean images from the second part to train ADANI. Quantitative results are listed in Table \ref{table_real}. As observed, ADANI achieves PSNR and SSIM comparable to other fully-supervised networks. Visual results are presented in Figure \ref{fig_sidd}. Since the noise in SIDD data is spatially correlated, which 
violates the assumption of N2V, it fails to remove this noises, unlike our proposed method.

\begin{figure}
	\centering
	\subfigure[Clean{\color{white} aaaaaa}  SSIM, PSNR]{\includegraphics[width=0.75in]{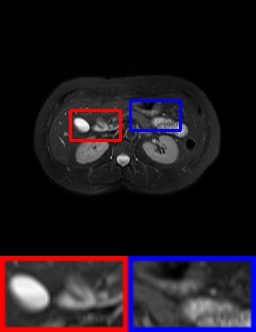}}
	\subfigure[Input {\color{white} aaaaaa} 0.129, 19.67  ]{\includegraphics[width=0.75in]{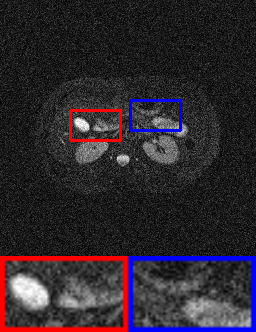}}
	\subfigure[U-Net {\color{white} aaaaa} 0.938, 37.02]{\includegraphics[width=0.75in]{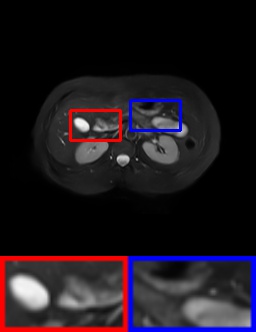}}
	\subfigure[N2B {\color{white} aaaaaa} 0.928, 35.86]{\includegraphics[width=0.75in]{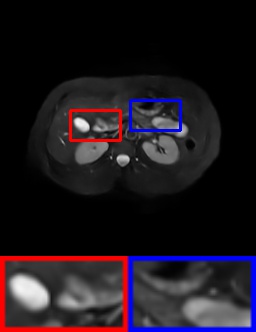}}

	\caption{MRI denoising example. }
	\label{fig_mri}
\end{figure}

\begin{figure}[t]
	\centering

	\subfigure[ {Clean $|$ SSIM, PSNR}]{\includegraphics[width=1.4in]{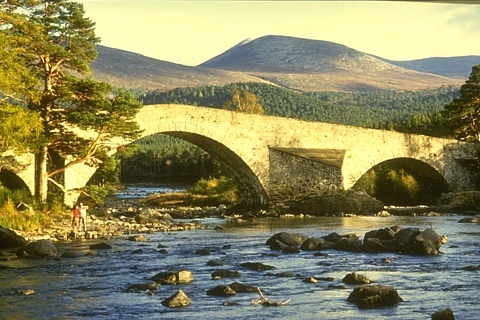}}
	\subfigure[ {Noise $|$ 0.746, 18.46}]{\includegraphics[width=1.4in]{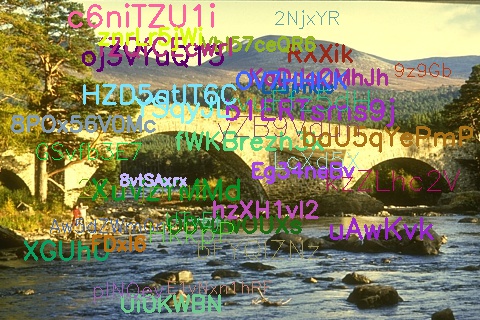}}\\
	\vspace{-0.1in}
	\subfigure[ {U-Net $|$ 0.957, 31.31}]{\includegraphics[width=1.4in]{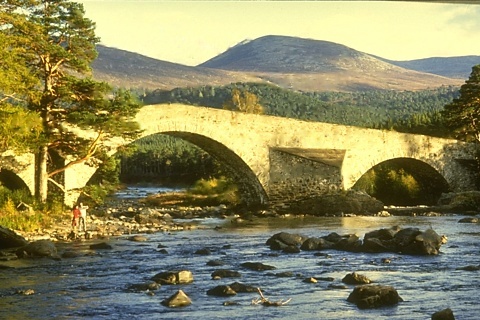}}
	\subfigure[ {Our $|$ 0.937, 28.95}]{\includegraphics[width=1.4in]{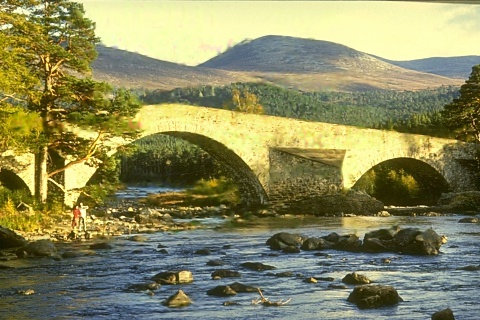}}
	
	\caption{Example results for text inpainting, $p=0.15$.}
	\label{fig_text}
\end{figure} 

\subsection{MRI denoising}

Magnetic resonance imaging (MRI) is a non-invasive medical imaging technology, which can provide high-resolution images of human tissues and organs. The quality of MR image, however, is easily degraded by noise during image acquisition. The noise in MR images follows the Rician distribution, which is much more complex than traditional additive noise. Here, we show the ability of ADANI for MR image denoising. We conduct experiments on the liver images of the CHAOS\footnote{https://chaos.grand-challenge.org/} dataset. We randomly sample half of the clean images from the training set, and each image is degraded by a different level ($1\% - 13\%$ of maximum intensity) of Rician noise \cite{ran2019denoising}. The remaining clean images belong to $D^{clean}$. The 600 images in test set are damaged by noise at a level of $8\%$. ADANI is compared with the fully-supervised U-Net. Results are shown in Figure \ref{fig_mri}. Our denoising network cleanly removes noise and restores high-quality images. In addition, U-Net gives an average of 0.905/35.63 dB in terms of SSIM and PSNR, slightly better than the ADANI of 0.896/34.77 dB.

\subsection{Blind image inpainting}
ADANI can be applied to other image restoration tasks. Here, we show the application of ADANI in image inpainting. Similar to denoising, for image inpainting, ADANI requires neither a priori of the image degradation process nor the paired data. We use the clean images in subsection \ref{synthetic} to synthesize text-degraded data. This degradation contains a variety of random strings, which can be random font sizes, random colors, and random locations. Each pixel in the training samples is degraded with a variable probability $p\in \left(0,0.3\right]$. For test images, p is fixed to 0.15. ADANI is compared with a fully supervised U-Net. Subjective comparisons are presented in Figure \ref{fig_text}. ADANI gives $0.942/30.78 dB$ in terms of SSIM and PSNR for the BSD300 test set, close to the U-Net of $0.964/33.75 dB$.

\begin{table}[t]
	\caption{Ablation study on the effect of $\alpha \mathcal{L}_{gradient}$ and $\gamma \mathcal{L}_{logit}$.}
	\centering
	\scriptsize
	\begin{tabular}  {|c|p{0.05\columnwidth}|p{0.09\columnwidth}|p{0.11\columnwidth}|p{0.12\columnwidth}|p{0.11\columnwidth}|p{0.11\columnwidth}|}
		\hline
		\multicolumn{2}{|c|}{} & $\alpha=0$  $\gamma=0$&$\alpha=10$  $\gamma=0$&$\alpha=100$  $\gamma=0$&$\alpha=z^r_1$  $\gamma=0$&$\alpha=z^r_1$  $\gamma=0.1$\\
		\hline
		
		Speckle  &SSIM&0.819&0.856&0.830&0.868&\textbf{0.872}\\
		\cline{2-7}
		($v=0.1$)&PSNR&27.37&29.06&28.15&29.70&\textbf{29.96}\\
		\hline

	\end{tabular}
	\label{table_ablation}
\end{table}

\subsection{Ablation study}

The core of ADANI is the noise similarity loss $\alpha \mathcal{L}_{gradient}$ and logit consistency loss $\gamma \mathcal{L}_{logit}$ in Eq.(\ref{eq.L_all_final}). To verify the importance of the proposed adaptive noise imitation strategy, we compare ADANI with its 4 variants. In the first one, we set the hyperparameter $\gamma$ for $\mathcal{L}_{logit}$ to 0. For the remaining three examples, we further set $\alpha$ to the value sampled from $\{0,10,100 \}$ instead of the dynamic $z^r_1$. Similar to subsection \ref{synthetic}, we conduct the experiments over the speckle noise dataset with $v \in \left(0, 0.2\right]$. The comparison is reported in Table \ref{table_ablation}. $\mathcal{L}_{logit}$ is conducive to the generation of diverse noises, leading to the improvement of the denoising performance of our method. On the other hand, setting the hyperparameter $\alpha$ for $\mathcal{L}_{gradient}$ to 10 or 100 compromises our denoising network. This is because the importance of all noisy images is the same, which makes it difficult for ADANI to distinguish noise from the image edges in the gradient map ${\nabla }x^r$. If both $\alpha$ and $\gamma$ are 0, our ADANI turns into a normal GAN. Since the ambiguous GAN loss $\mathcal{L}_{GAN}$ cannot permit high-quality noise generation, the denoising results are poor. In contrast, our adaptive noise imitation strategy achieves high-quality noise synthesis and denoising.

\section{Conclusion}
We proposed a novel adaptive noise imitation (ADANI) algorithm, which enables the training of denoising CNNs without pre-collected paired data. ADANI generates new noisy data for learning denoising by observing its unpaired noisy/clean input. The noisy data produced by ADANI is visually and statistically similar to the real one, so that it achieves encouraging denoising performance. We demonstrate the effectiveness and wide applicability of ADANI over multiple denoising and image restoration tasks. Since ADANI does not require pre-collected paired data and pre-defined image degradation processes, it is a promising solution in many practical applications.

{\small
	\bibliographystyle{ieee}
	\bibliography{egbib}
}

\end{document}